\documentclass[twocolumn, showpacs]{revtex4}
\usepackage{hyperref}
\usepackage{amsmath}
\usepackage{color}
\usepackage[dvips]{graphicx}
\usepackage{amsmath, amssymb}

\begin{document}
\title{Terahertz Time-Domain Magnetospectroscopy of a High-Mobility \\ Two-Dimensional Electron Gas}
\author{Xiangfeng Wang}
\author{David J. Hilton}
\author{Lei Ren}
\author{Daniel M. Mittleman}
\author{Junichiro Kono}
\email{kono@rice.edu}
\address{Department of Electrical and Computer Engineering, Rice University, Houston, Texas 77005, USA}
\author{John L.~Reno}
\address{Sandia National Laboratories, P.~O.~Box 5800, Albuquerque, New Mexico 87185, USA}
\date{\today}

\begin{abstract}
We have observed cyclotron resonance in a high-mobility GaAs/AlGaAs two-dimensional electron gas by using the techniques of terahertz time-domain spectroscopy combined with magnetic fields.  From this, we calculate the real and imaginary parts of the diagonal elements of the magnetoconductivity tensor, which in turn allows us to extract the concentration, effective mass, and scattering time of the electrons in the sample.  We demonstrate the utility of ultrafast terahertz spectroscopy, which can recover the true linewidth of cyclotron resonance in a high-mobility ($>{10}^{6}\,\mathrm{cm^{2}\,V^{-1}\,s^{-1}}$) sample without being affected by the saturation effect.
\end{abstract}

\maketitle


Quantum coherence is an important ingredient in modern condensed matter physics as well as in emerging technologies.
The creation and manipulation of a coherent superposition of two or multiple quantum states is the subject of many current studies.
An ultrahigh-mobility two-dimensional electron gas (2DEG) offers an ideal system for studying novel quantum coherent phenomena in a clean, solid-state environment.
In particular, when one applies a magnetic field perpendicular to the 2DEG, the density of states splits into Landau levels, making a fully-tunable, atomic-like system.
In addition, a variety of phenomena that occur in the 2DEG arising from
carrier interactions, confinement, and disorder
can make quantum coherent effects even more exotic than in atomic or molecular
systems.  However, there has been little success in performing coherent spectroscopy of Landau-quantized 2DEGs, although there is a long history of cyclotron resonance (CR) studies of 2DEGs using Fourier-transform infrared (FTIR) spectroscopy~\cite{PalikFurdyna70RPP,Mavroides72MO,McCombeWagner75Review1,PetrouMcCombe91LLS,Nicholas94HOS,Konobook}.


Terahertz (THz) time-domain magnetospectroscopy~\cite{Crooker02RSI}, which combines conventional
THz time-domain spectroscopy (THz-TDS) with a high magnetic field, has a number of inherent advantages compared to traditional FTIR techniques.  THz-TDS directly measures both the amplitude and phase of the electric field $E\bigl(t\bigr)$ and allows for the simultaneous determination of the real and imaginary parts of the conductivity without using Kramers-Kronig techniques.
Additionally, use of a temporally-gated detection scheme, common to THz-TDS techniques,
significantly suppresses background thermal noise and results in an enhanced signal-to-noise ratio~\cite{Nuss9807,
SensingWithTerahertzRadiation}.

THz-TDS was used earlier~\cite{Some1994} to observe CR in relatively low-mobility ($\mu_e$ = 2.7 $\times$ 10$^5$~cm$^2$V$^{-1}$s$^{-1}$) 2DEG samples.  In addition, THz-TDS has been successfully employed to study quantum coherent phenomena in a wide range of systems, including the rotational transitions of N$_2$O molecules~\cite{Harde1991}, intersubband transitions in semiconductor quantum wells~\cite{Heyman}, and surface plasmons propagating on metal-film hole arrays~\cite{Du2004}.


Here, we report the observation of long-lived, magnetic-field-dependent coherent oscillations in a \emph{high-mobility} GaAs/AlGaAs 2DEG in a perpendicular magnetic field.  We explain our observations in terms of a coherent superposition created by the incident THz pulse between the lowest unfilled Landau level and the highest filled Landau level.  In addition, we determine elements of the complex magnetoconductivity tensor $\tilde\sigma$
as a function of both frequency $\nu$ and magnetic field $B$, which in turn allows us to determine the cyclotron frequency $\nu_c$, effective mass $m^*$, and cyclotron resonance linewidth $\Delta\nu_c$ (or the scattering time $\tau=1/\Delta\nu_c$) as a function of $B$.  Finally, we show that THz-TDS can overcome the ``saturation effect''~\cite{Chou1988,Studenikin2005} that often prevents FTIR-based techniques from determining the true linewithds of CR in high-mobility ($\mu_e > 10^6~{\rm cm}^2{\rm V}^{-1}{\rm s}^{-1}$) 2DEGs.  We successfully measured the linewidth as a function of temperature and magnetic field, which will be reported in detail elsewhere. 

Broadband, ultrashort THz pulses were generated and detected using a standard photoconductive antenna-receiver
setup~\cite{Nuss9807,SensingWithTerahertzRadiation}.  Time delay was provided by an oscillating retro-reflector operating at 3~Hz; each data set was the averaged result of $\sim 800$ scans.  We used an Oxford superconducting magnet (SM-4000-10T) to produce fields ranging from 0 to 1.4~T and temperatures from 1.5~K to 300~K.  The sample studied in this experiment was a modulation-doped GaAs/AlGaAs single quantum well with an electron concentration of $n_e = 2.0 \times {10}^{11}\,
\mathrm{cm}^{-2}$ and mobility of $\mu_e = 3.7 \times {10}^{6}\,
\mathrm{cm^{2}\,V^{-1}\,s^{-1}}$ at 4.2~K, determined through Shubnikov-de Haas and DC conductivity measurements.


\begin{figure}
\includegraphics[scale=0.63]{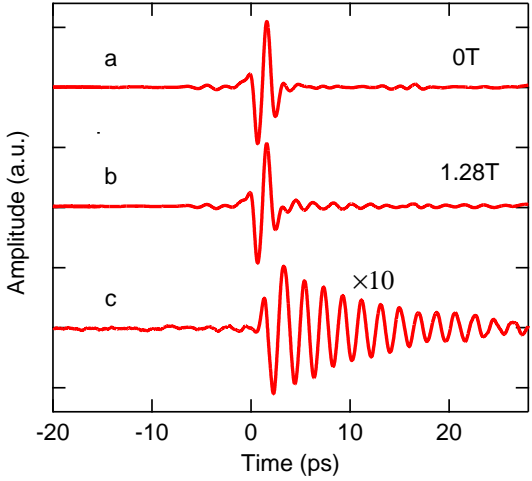}
\caption{(color online) THz waveforms transmitted through a high-mobility 2DEG at 0~T (a) and at 1.28~T (b) at 2 K.  The cyclotron oscillations induced by the magnetic field (c) are isolated by subtracting (a) from (b). \label{WangF1}}
\end{figure}

\begin{figure}
\centerline{\includegraphics[scale=0.45]{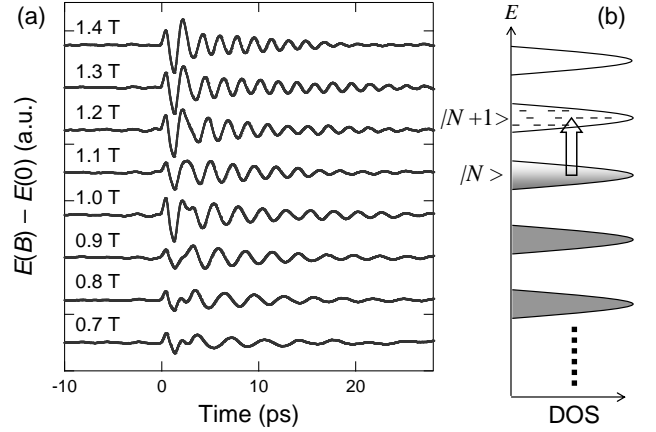}}
\caption{(a) Time-domain cyclotron oscillations in a high-mobility 2DEG from 0.7~T to 1.4~T at 2~K.  Traces are vertically offset for clarity. (b) A Landau-quantized 2DEG. DOS: density of states.\label{wangF2}}
\end{figure}

In this experiment, the THz waveform is measured after transmission through the sample in a $B$ field from 0~T to 1.4~T.  Figure \ref{WangF1} plots these waveforms at 0~T [trace~(a)] and at a finite $B$ (1.28 T) [trace~(b)].  Trace~(c) is the difference between the transmitted THz electric field at 1.28~T and 0~T highlighting the change to the THz transmission due to the $B$ field (data enlarged 10 times), which shows the $B$-induced oscillations of the electric field of the THz pulse.  We verify that the observed oscillations originate from the 2DEG and not a $B$-dependence of any of the optics in the experiment by first measuring the $B$-dependent THz transmission in the absence of the 2DEG in an otherwise identical configuration.  Figure~\ref{wangF2}(a) shows similar oscillations
induced in the transmitted THz waveform from 0.7~T to 1.4~T whose frequency
and decay time vary with $B$.

\begin{figure}
\centerline{\includegraphics[scale=0.5]{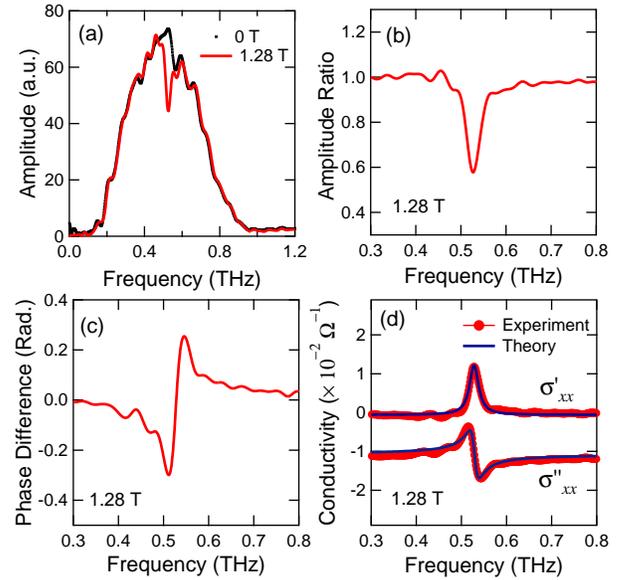}} \caption{(color
online) (a) Amplitude of the transformed electric fields at 0~T and 1.28~T. (b) Magnitude of the complex transmission coefficient at 1.28~T.  (c) Phase of the transmission coefficient.  (d) Real ($\sigma_{xx}'$) and imaginary ($\sigma''_{xx}$) parts of the magnetoconductivity tensor element $\sigma_{xx}$ at 1.28~T. The $\sigma''_{xx}$ trace is vertically offset. \label{wangF3}}
\end{figure}

Figure \ref{wangF3}(a) shows the amplitude of the Fourier-transformed electric fields at 0~T and 1.28~T.  A $B$-field-induced absorption, or a dip, is clearly seen in the 1.28~T spectrum.  Figure \ref{wangF3}(b) shows the magnitude of the complex transmission coefficient ($T = \bigl\lvert T\bigr\rvert e^{i \phi}$) at 1.28~T, while Fig.~\ref{wangF3}(c) shows the phase, $\phi$. In our experiment, the THz pulse is linearly polarized ($\hat x$), and we detect only one polarization component $(\hat x$) after transmission through the sample.  As a result, our measurement is dependent on the corresponding diagonal element of the magnetoconductivity tensor, $ \sigma_{xx}$~\cite{PalikFurdyna70RPP}.  In order to extract the conductivity from the complex transmission coefficient, we model this sample as a thin conducting sheet on a thick substrate with an index of refraction, $n$.  In this approximation, the ratio of the Fourier transform of the waveforms at a finite $B$ field, $E\bigl(\nu, B\bigr)$, to the zero-field transmitted spectrum, $E\bigl(\nu, 0\bigr)$, is given by:
\begin{equation}
T_{xx}\bigl(\nu, B\bigr)={{E\bigl(\nu, B\bigr)}
\over{E\bigl(\nu, 0\bigr)}} = {2Y \over {2Y +
\sigma_{xx}\bigl(\nu, B\bigr)} },
\end{equation}
where $Y = n/Z_0$ is the admittance of the GaAs substrate and $Z_0 = 377~\mathrm{\Omega}$ is the impedance of free space~\cite{Nuss9807}.  Figure~\ref{wangF3}(d) highlights real ($\sigma_{xx}'$) and imaginary ($ \sigma''_{xx}$) parts of the extracted conductivity tensor element at 1.28~T.  Due to the rotational symmetry of the system perpendicular to the plane of the 2DEG, we would expect the same
result in the case of input polarization and detection both along
$\hat y$, i.e., $\sigma_{xx} = \sigma_{yy}$~\cite{PalikFurdyna70RPP}.

We determine the cyclotron frequency $\nu_c$ (s$^{-1}$), the cyclotron resonance linewidth $\Delta\nu_c = 1/\tau$ (s$^{-1}$), and the magnitude of the conductivity $\sigma_0$ ($\Omega^{-1}$) by fitting the results shown in Figs.~\ref{wangF3}(c) and 3(d).  Both the real ($\sigma_{xx}'$) and imaginary ($
\sigma''_{xx}$) parts of the magnetoconductivity tensor element are fit by
%
\begin{equation}\label{eq:ACDrude}
\sigma_{xx} = \sigma_{xx}' +i \sigma''_{xx} = \frac{\sigma_0}{1+{2\pi}i\bigl(\nu-\nu_{c}\bigr)\tau}.
\end{equation}
%
A representative fit at 1.28~T is shown in Fig.~\ref{wangF3}(d), using $\sigma_0 = 0.0126~\Omega^{-1}$,
$\nu_{c}$ = 0.529~THz, and $\tau$ = 15.6~ps.

An applied $B$ field perpendicular to the 2DEG results in the
formation of discrete Landau levels [see Fig.~\ref{wangF2}(b)] with an energy separation, $\Delta E$, between the $\bigl\lvert N\bigr\rangle$ and $\bigr\lvert N+1\rangle$ levels given by
\begin{equation}\label{eq:LLE}
\Delta E = \hbar\frac{ e B}{m^{*}} = h\nu_{c}
\end{equation}
where $e$ is the electron charge and $h$ is Planck's constant.  An incident THz wave with a photon energy equal to this separation coherently creates a superposition state between the highest filled Landau level, $\bigl\lvert N\bigl\rangle$, and the lowest unfilled Landau level, $\bigl\lvert
N+1\bigl\rangle$, as shown in Fig.~\ref{wangF2}(b).  This results in an atomic-like two-level system; all other Landau levels are either completely filled or completely empty (as long as the THz field is sufficiently weak as in our experiment) and do not affect the transmission of the THz pulse.  The observed damped oscillations in our experimental data can thus be viewed as the free induction decay~\cite{Allen1987} of such coherently coupled Landau levels.

Using the extracted value of $\nu_c$ and Eq.~\eqref{eq:LLE}, we obtain a value of the effective mass of $m^{*} = 0.0676\,m_{0}$, where $m_{0} = 9.11 \times 10^{-31}$~kg is the free electron mass.  Also, using the extracted values of $\sigma_0$ and $\tau$, we can determine the value of the electron concentration of $n_e
= 1.95 \times 10^{11}\,\mathrm{cm^{-2}}$, which is consistent with
the concentration obtained from transport measurements ($2.0
\times 10^{11}\,\mathrm{cm^{-2}}$).
Finally, the extracted linewidth is a measure of the scattering mechanisms present in the sample at this temperature and $B$ field.  We have systematically studied the temperature and $B$ dependence of $\Delta\nu_c$, which would allow us to elucidate a detailed theoretical understanding of the physical origins of this linewidth and will be reported in detail eslewhere.


In high-mobility samples, the apparent linewidth determined by FTIR measurements is much larger than the true linewidth, a phenomenon commonly referred to as the ``saturation effect''~\cite{Chou1988,Studenikin2005}.  This results from the decrease in \emph{detectable} transmission of the THz radiation over a broad spectral range; as the conductivity increases with either increasing carrier concentration or mobility, a spectral region exists with a finite width where, effectively, no transmission is
permitted.  The lack of a phase sensitive detection scheme in traditional FTIR techniques makes the direct determination of the complex conductivity in this situation difficult; the broadened linewidth in this case could result either from the increase in mobility or concentration.  Because of this saturation effect, almost no systematic linewidth studies exist for high-mobility 2DEG samples ($>10^6\,\mathrm{cm^2\,V^{-1}\,s^{-1}}$).

In order to overcome the lack of phase sensitive detection in FTIR measurements, different methods have been proposed.  For example, measurement of the transmission coefficient of a 2DEG over a broad spectral range will permit the use of Kramers-Kronig techniques to calculate the phase at THz frequencies and determine the complex conductivity~\cite{Kuzmany1998}.  A second alternate method for determining the complex conductivity \emph{assumes} a Drude form for $\tilde \sigma$ and fits this
to the measured intensity transmission coefficient; lack of a direct phase measurement makes this an ambiguous determination of $n_e$ and $\tau$~\cite{Chou1988}.

THz-TDS allows for the \emph{direct} determination of the full complex conductivity of the sample without resort to Kramers-Kronig techniques and without an \emph{a priori} assumption of the lineshape function.  The increased signal-to-noise ratio inherent to the gated detection scheme allows for the determination of the lower transmitted THz signals that result from high-mobility and high-concentration 2DEGs.  Second, the additional spectroscopic information determined from the phase sensitive measurement removes the ambiguity between $n_e$ and $\tau$.  As a result, no assumption of lineshape is necessary in order to calculate the full complex conductivity.  Employing this technique allows for the simultaneous determination of both $n_e$ and $\tau$ from the measurement of the transmitted THz waveform electric field.

In summary, we have observed time-domain cyclotron resonance oscillations in a GaAs/AlGaAs 2DEG, which can be modeled as the decay of the coherent superposition of two coupled Landau levels induced by the incident THz pulse. The real and imaginary parts of the conductivity are determined simultaneously at different magnetic fields without using Kramers-Kronig analysis. We show that our THz technique has many advantages for doing cyclotron resonance measurements, especially for high-mobility samples.

This work was supported by the National Science Foundation (through Grant Nos.~DMR-0134058 and DMR-0325474). Sandia is a multiprogram laboratory operated by Sandia Corporation, a Lockheed Martin Company, for the National Nuclear Security Administration under
Contract DE-AC04-94AL85000.


\end{document}